\documentclass[useASM]{mn2e}
\usepackage{graphics,graphicx}
\usepackage{subfigure}
\usepackage{epsfig}
\usepackage{appendix}
\usepackage{lineno}
\usepackage{url}
\usepackage{bigstrut,bigdelim,multirow}

\begin{document}

\title[The properties of Short GRBs with EE]
{The properties of prompt emission in short GRBs with extended emission observed by {\em Fermi}/GBM}
\author[Lan et al.]
{Lin Lan$^{1}$, Rui-Jing Lu$^{1}$, Hou-Jun L\"{u}$^{1}$\thanks{E-mail: lhj@gxu.edu.cn}, Jun
Shen$^{1}$, Jared Rice$^{2}$,
Long Li$^{1}$, and En-Wei Liang$^{1}$\\
$^1$Guangxi Key Laboratory for Relativistic Astrophysics, School of Physical Science and
 Technology, Guangxi University, Nanning 530004, China\\
 $^2$Department of Physics, Texas State University, San Marcos, TX 78666, USA\\}
 \maketitle

\label{firstpage}
\begin{abstract}
Short GRBs with extended emission (EE) that are composed initially of a short-hard spike and
followed by a long-lasting EE, are thought to be classified as a subsection of short GRBs. The
narrow energy band available during the {\em Swift} era combined with a lack of spectral
information prevented discovery of the intrinsic properties of those events. In this paper, we
performed a systematic search of short GRBs with EE by using all available {\em Fermi}/GBM data.
The search identified 26 GBM-detected short GRBs with EE that are similar to GRB 060614 observed by
{\em Swift}/BAT. We focus on investigating the spectral and temporal properties for both the hard
spike and the EE components of all 26 GRBs, and explore differences and possible correlations
between them. We find that while the peak energy ($E_{\rm p}$) of the hard spikes is a little bit
harder than that of the EE, but their fluences are comparable. The harder $E_{\rm p}$ seems to
correspond to a larger fluence and peak flux with a large scatter for both the hard spike and EE
components. Moreover, the $E_{\rm p}$ of both the hard spikes and EE are compared to other short
GRBs. Finally, we also compare the properties of GRB 170817A with those short GRBs with EE and find
no significant statistical differences between them. We find that GRB 170817A has the lowest
$E_{\rm p}$, likely because it was off-axis.

\end{abstract}

\section{Introduction}
Phenomenologically, gamma-ray bursts (GRBs) have been generally divided into ``long soft" and
``short hard" classes based on the observed bimodal distribution in duration and hardness ratio
(Kouveliotou et al. 1993). The progenitors of long GRBs likely originate from the core collapse of
a massive star, e.g. via observations of associated supernovae (Narayan et al. 1992; Woosley 1993;
Galama et al. 1998; Hjorth et al. 2003; Stanek et al. 2003; Woosley \& Bloom 2006), and the
progenitors of short GRBs are likely the coalescence of two compact objects, i.e. neutron star -
neutron star (NS-NS) or neutron star - black hole (NS-BH) systems (Paczynski 1986; 1991 Eichler et
al. 1989).

Within the short GRB class, there is a subsection of bursts that is characterized by a short/hard
spike (with a duration $\sim$5 s) followed by a series of soft gamma-ray pulses with a much longer
duration (called extended emission; Norris \& Bonnell 2006; Troja et al. 2008; Perley et al. 2009).
Since the discovery of the first clear evidence of extended emission (EE) in GRB 060614 (Gehrels et
al. 2006; Gal-Yam et al. 2006; Fynbo et al. 2006; Della Valle et al. 2006), there has been an
extensive search for more of these types of events are in both the {\em Swift} (Zhang et al. 2009;
Norris et al 2010; Sakamoto et al. 2011) and {\em Fermi} eras (Kaneko et al. 2015).

From the theoretical point of view, a number of different models have been proposed to interpret
short GRBs with EE. For instance, the EE could be the product of an accretion disc around a
magnetar undergoing magnetic propellering (Metzger et al. 2008; see also Zhang \& Dai 2008, 2009;
Piro \& Ott 2011; Gompertz et al. 2013; Bernardini et al. 2014; Gibson et al. 2017), the magnetic
dipole spin-down of a magnetar (Dai \& Lu 1998a,b; Zhang \& M\'esz\'aros 2001; Fan \& Xu 2006;
Bucciantini et al. 2012; Rowlinson et al. 2013; L\"{u} et al. 2015), a two-jet solution (Barkov \&
Pozanenko 2011), r-process heating of the accretion disc (Metzger et al. 2010), or magnetic
reconnection and turbulence (Zhang \& Yan 2011). Liu et al. (2012) suggested that the short GRBs
with EE may arise from radial angular momentum transfer in the disk and the magnetic barrier around
the black hole. L\"{u} et al. (2015) proposed that EE components detected in the BAT band could be
simply the internal plateau emission when that emission is bright and hard enough. In any case, the
rapid variability of this EE strongly suggests that it results from ongoing central engine activity
(Perley et al. 2009; Metzger et al. 2011).

The central engine and radiation mechanism of short GRBs with EE remain open questions, but the
intrinsic spectra of both the hard spike and EE components may provide some important clues for
understanding these questions. In the {\em Swift} era, some systematic analyses of the spectral
properties of short GRBs with EE show that the EE component is softer than the hard spike
(Villasenor et al. 2005; Norris \& Bonnell 2006; Troja et al. 2008; Perley et al. 2009). However,
the spectra of both the initial hard spike and subsequent EE components are well-fitted by a
power-law model, which can not reflect the intrinsic properties of the spectra. Kaneko et al.
(2015) performed a systematic comparison of the short GRBs with EE observed by {\em Swift}/BAT with
those observed by {\em Fermi}/GBM. However, they only considered the NaI detectors of GBM in their
spectral analysis and ignored the contributions from the BGO detectors. This may be the reason why
most of the short GRBs with EE can be fitted with a power-law model in Kaneko et al (2015).

More interestingly, the short GRB 170817A associated with the gravitational wave event (GW 170817)
from double a NS merger was recently detected by {\em Fermi}/GBM (Abbott et al. 2017; Goldstein et
al. 2017), and was found to have a prompt emission that is composed of a hard spike of 0.8 s and a
weak tail of up to 2.3 s (Goldstein et al. 2017; Zhang et al. 2018; Pozanenko et al. 2018). An
interesting question is whether or not GRB 170817A differs in its spectral properties from other
typical short GRBs with EE.

Although the short GRBs with EE have been investigated from both statistical and theoretical
analyses, it remains unclear whether the initial hard spike differs in spectral properties from the
subsequent EE component. What are the relationships between the hard spike and subsequent EE
component? Do the hard spike and subsequent EE component share the same physical origin? The aim of
this paper is to address these interesting questions through a systematic analysis of the {\em
Fermi}/GBM data by considering the contribution of the high-energy detector. Our data reduction and
sample selection are presented in \S2. Some comparisons between the hard spike and subsequent EE
component for our sample, as well as other typical short GRBs and GRB 170817A, are reported in \S3.
The conclusions and discussion are drawn in \S4.

\section{Data reduction and sample selection}
The {\em Fermi} satellite has operated for more than ten years, and provides unprecedented spectral
coverage over seven orders of magnitude in energy (from $\sim$8 keV to $\sim$300 GeV). There are
two instruments onboard the {\em Fermi} satellite. One is the Gamma-ray Burst Monitor (GBM; Meegan
et al. 2009), which has twelve sodium iodide (NaI) and two bismuth germanate (BGO) scintillation
detectors covering an energy band from 8 keV to 40 MeV. There are three types of data modes CTIME,
CSPEC, and TTE, which correspond to a time resolution of 64 ms, 1.024 s, and any bin size,
respectively (Paciesas et al. 2012). We select only the TTE data in our analysis due to including
individual photons arriving with time and energy tags\footnote{CTIME and CSPEC data are not used in
our analysis due to the fixed time resolution of 64 ms and 1.024 s, respectively.}, and any time
resolution bin size can be selected to perform the spectral and temporal analysis. The other
instrument onboard {\em Fermi} is the Large Area Telescope (LAT) with an energy coverage from 20
MeV to 300 GeV (Atwood et al. 2009). Here, we consider only GBM data for the temporal and spectral
analysis and ignore the contributions of LAT data because the physical origin of high-energy
photons remain an open question (e.g., originating from internal or external dissipation).

\subsection{Lightcurve extraction}
We obtain the original GBM data (from the twelve NaI and two BGO detectors) from the public science
support center at the official {\em Fermi} web
site\footnote{http://fermi.gsfc.nasa.gov/ssc/data/.}. We select the brightest detector in NaI and
BGO for our analysis because the brightest detector has the minimum angle between the incident
photon and the normal direction of the detector. Based on the standard \texttt{heasoft} tools
(version 6.19) and the {\em Fermi} \texttt{ScienceTools} (v10r0p5), we developed a {\em Python}
code to extract the energy-dependent lightcurves and time-dependent spectra using the spectral
source package $gtBurst$\footnote{http://sourceforge.net/projects/gtburst/.}. For more details of
lightcurve extraction with the Bayesian Block algorithm, please refer to our latest paper Lan et
al. (2018). Moreover, we calculated the duration of both the hard spike ($T_{\rm d, s}$) and
subsequent EE ($T_{\rm d, e}$), which are reported in Table 1.

\subsection{Sample selection criteria}
As of December 2018, we have extracted the lightcurves of more than 2400 GRBs which were detected
by {\em Fermi} /GBM. GRB 060614 was the first clear case of a short GRB with EE; its lightcurve of
prompt emission is composed of a short spike with $\sim$5 s and followed a longer soft emission
(Gehrels et al. 2006; Gal-Yam et al. 2006; Fynbo et al. 2006; Della Valle et al. 2006). We adopt
the properties of the lightcurve of GRB 060614 as the ``standard event" to search for in our
samples, with the following three criteria:
\begin{enumerate}
\item The duration of the initial hard spike is less than 5 seconds, and is followed by a
    longer soft emission lasting a few seconds to hundreds of seconds.
\item The signal-to-noise ratio (S/N) of the initial spike and EE components should be greater
    than 3$\sigma$.
\item The count rates remain below 30\%$\sim$40\% of the peak count rate for at least 50\% of
    the rest of the duration after the peak time until $T_0$+5\,s (see also Kaneko et al.
    2015).
\end{enumerate}
There are 26 GRBs that satisfy with our criteria up to December 2018. No redshifts are measured in
our sample. A comparison of our sample with that of Kaneko et al. (2015) shows only four
overlapping GRBs, which may be due to different sample selectrion criteria. An example lightcurve
from our sample is the GRB 161218B lightcurve shown in Figure \ref{fig:LC}.

\subsection{Spectrum Extraction and fitting}
\begin{figure*}
\begin{tabular}{lllll}
\multirow{3}{*}{ \includegraphics[angle=0,scale=0.40]{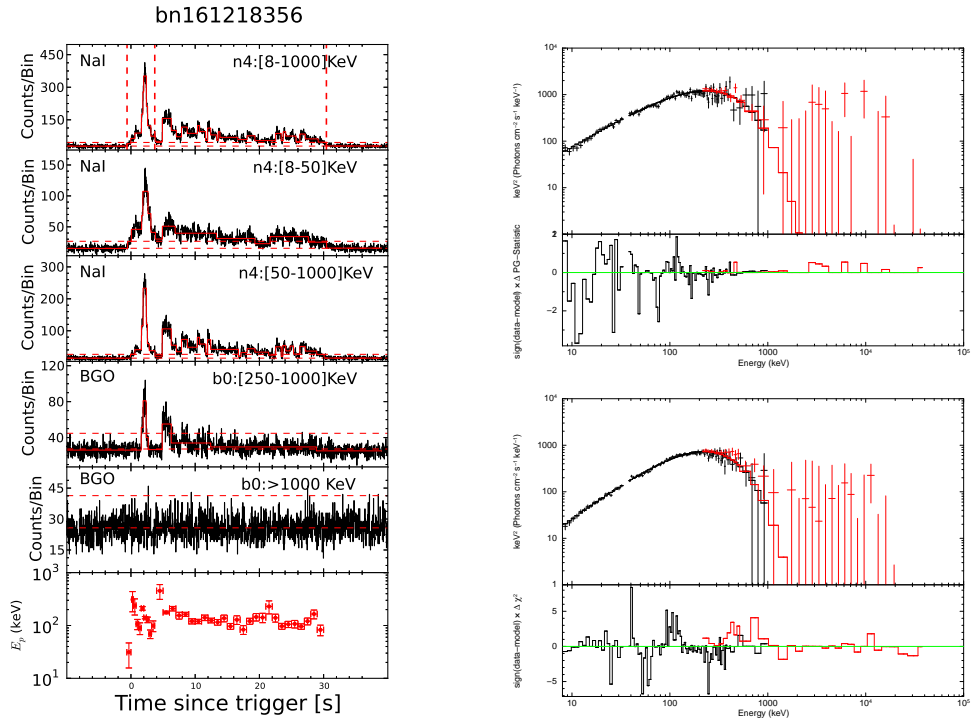}} \\
&\includegraphics[angle=270,scale=0.25]{f1b.eps} \\
&\includegraphics[angle=270,scale=0.25]{f1c.eps} \\
\end{tabular}
\hfill\center \caption{An example of GRB 161218 lightcurve and spectrum, together with our Bayesian
block analysis (red blocks in the left panel), time-integrated spectral fits for the hard spike and
subsequent EE (solid line in the right panel), and $E_{\rm p}$ evolution of the prompt emission
phase. The dashed horizontal lines in the left panel are a 3$\sigma$ signal over background
emission. The dashed vertical lines are the beginning, separation, and end of the hard spike and
EE, respectively.} \label{fig:LC}
\end{figure*}

We select two time intervals that are long before and far after the prompt emission as the
background, and subtract it from the burst phase using a polynomial function fit. Then, XSPEC is
used to perform time-integrated spectral fits for the initial hard spike, subsequent EE, as well as
the entire burst (see Figure \ref{fig:LC}). In order to determine evolutionary behavior during the
prompt emission phase, the time-resolved spectra are also required (see Figure.1). The statistic
$\chi^2$ is adopted to judge the goodness of the spectral fits. Moreover, the energy channels in
the vicinity of the iodine K-edge at 33.17 keV were excluded to better assess the quality of the
fit of the spectral models (Goldstein et al. 2012).

A Band function model is the prevailing model for doing the spectral fit (Band et al. 1993).
Alternatively, a cutoff power-law (CPL) or simple power-law (PL) model can be fit if the Band
function model is not a good enough fit to the data. They can be written as
\begin{eqnarray}
N_{\rm CPL}(E) = A\cdot E^{-\alpha} {\rm exp}(-\frac{E}{E_{\rm p}}),
\end{eqnarray}
\begin{eqnarray}
N_{\rm PL}(E) = A\cdot E^{-\alpha},
\end{eqnarray}
where $A$ is the normalization of the spectrum, and $\alpha$ and $E_{\rm p}$ are the low-energy
photon spectral index and peak energy, respectively. On the other hand, we also attempt to take
into account a black body (BB) model or multi-component superposition models (e.g., BB+Band,
BB+CPL, and BB+PL) to fit the spectra of both the initial spike and EE, but they do not
significantly improve the goodness and they contain more free parameters compared to the CPL model.
Thus the CPL model is the optimal selection for both the hard spike and EE in our sample, except
the hard spike component of GRB 081215 (bn081215784) that can be fit with a Band function. An
example of spectral fitting in our sample is shown in Figure \ref{fig:LC}, and the spectral
parameters derived from our fits are reported in Table 1.

\section{Results}
\subsection{Statistical comparisons with the hard spike and EE}
As early as the Compton Gamma-Ray Observatory (CGRO) era, Ford et al. (1995) first found there are
two different $E_{\rm p}$-evolution patterns (i.e., hard-to-soft and intensity tracking) in the
prompt emission phase of long GRBs by performing a comprehensive analysis of Burst And Transient
Source Experiment (BATSE) data. After that, it was revisited by many other authors who obtained
similar results with those in Ford et al. (1995) (e.g., Liang \& Kargatis 1996; Borgonovo \& Ryde
2001; Kaneko et al. 2006; Lu et al. 2012). Here, we also analyze the time-resolved spectrum of both
the hard spike and subsequent EE for twenty GRBs of our sample\footnote{There are 26 short GRBs
with EE in our sample, but only 20 GRBs that have enough photons for an analysis of the
time-resolved spectrum.}. We find that the $E_{\rm p}$-evolution of five GRBs in our sample follow
the hard-to-soft pattern, twelve GRBs have intensity tracking, and the $E_{\rm p}$ of three GRBs
does not evolve significantly. We also derive the bolometric fluence in the 8 keV -- 40 MeV band
for both the hard spike ($S_{\rm \gamma, s}$) and EE ($S_{\rm \gamma, e}$), as well as the peak
flux. Moreover, we compare the peak energy of the hard spike ($E_{\rm p,s}$) with that of the EE
component ($E_{\rm p,e}$) by analyzing time-integrated spectral fits.

\begin{figure*}
\includegraphics[angle=0,scale=0.5]{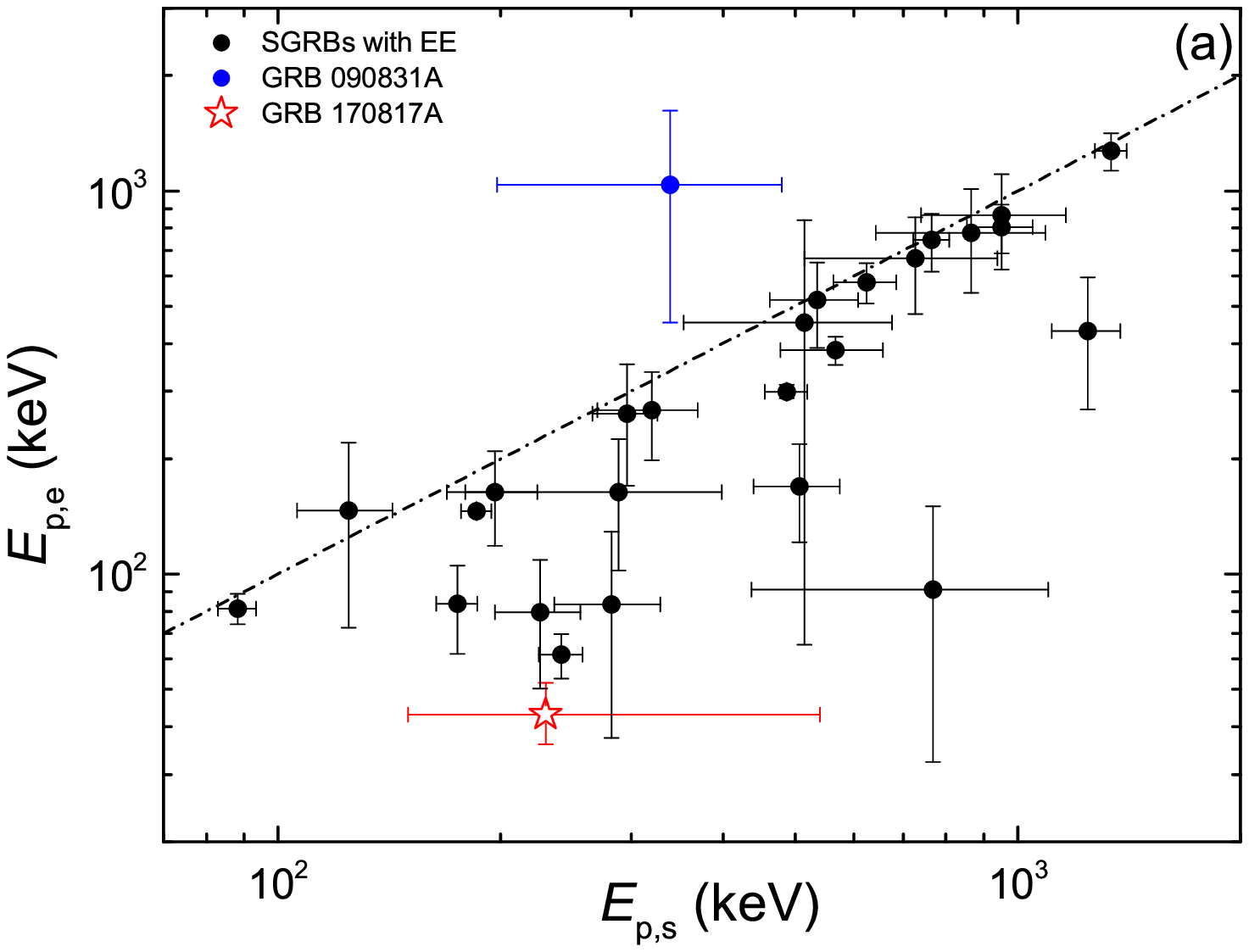}
\includegraphics[angle=0,scale=0.5]{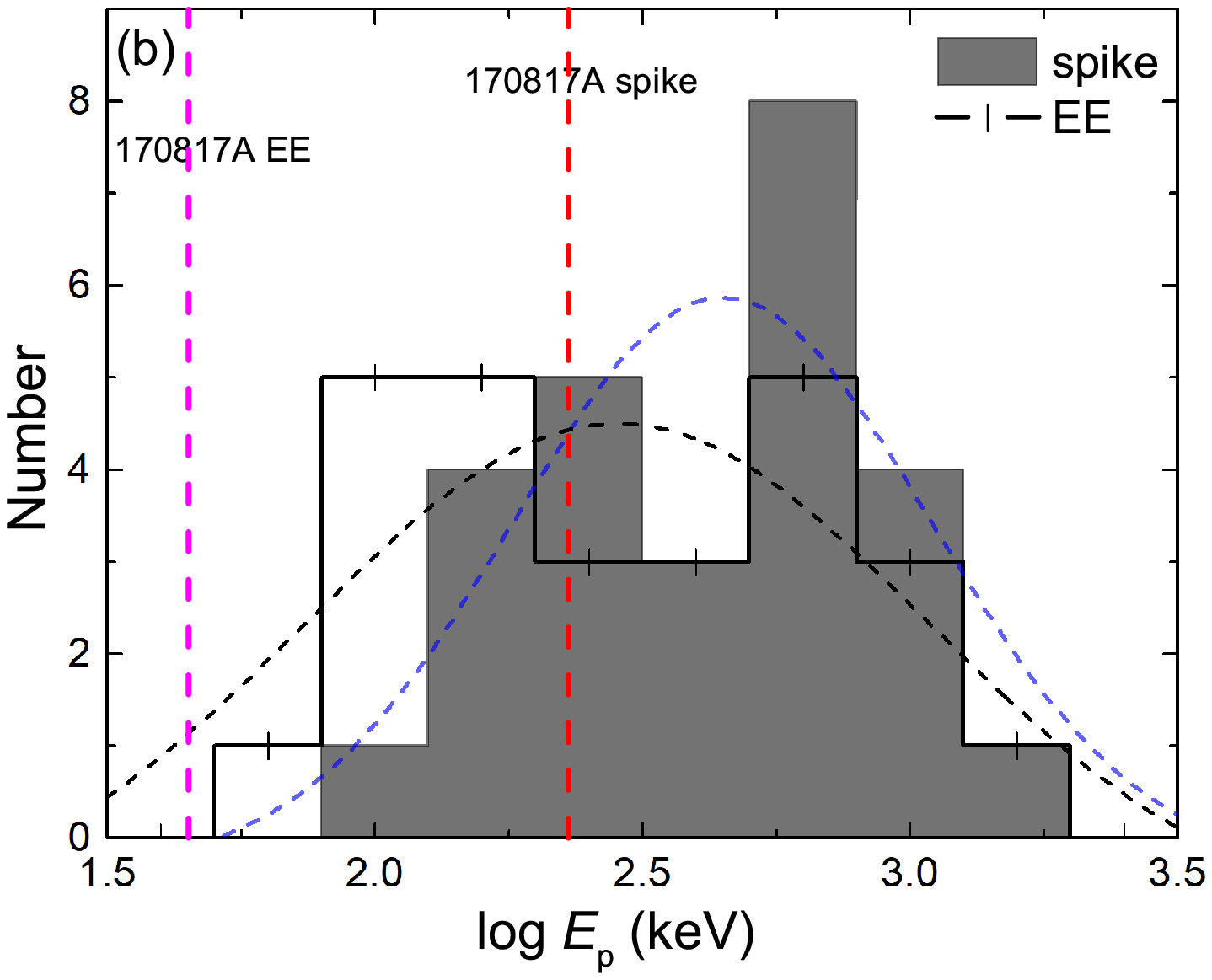}
\includegraphics[angle=0,scale=0.5]{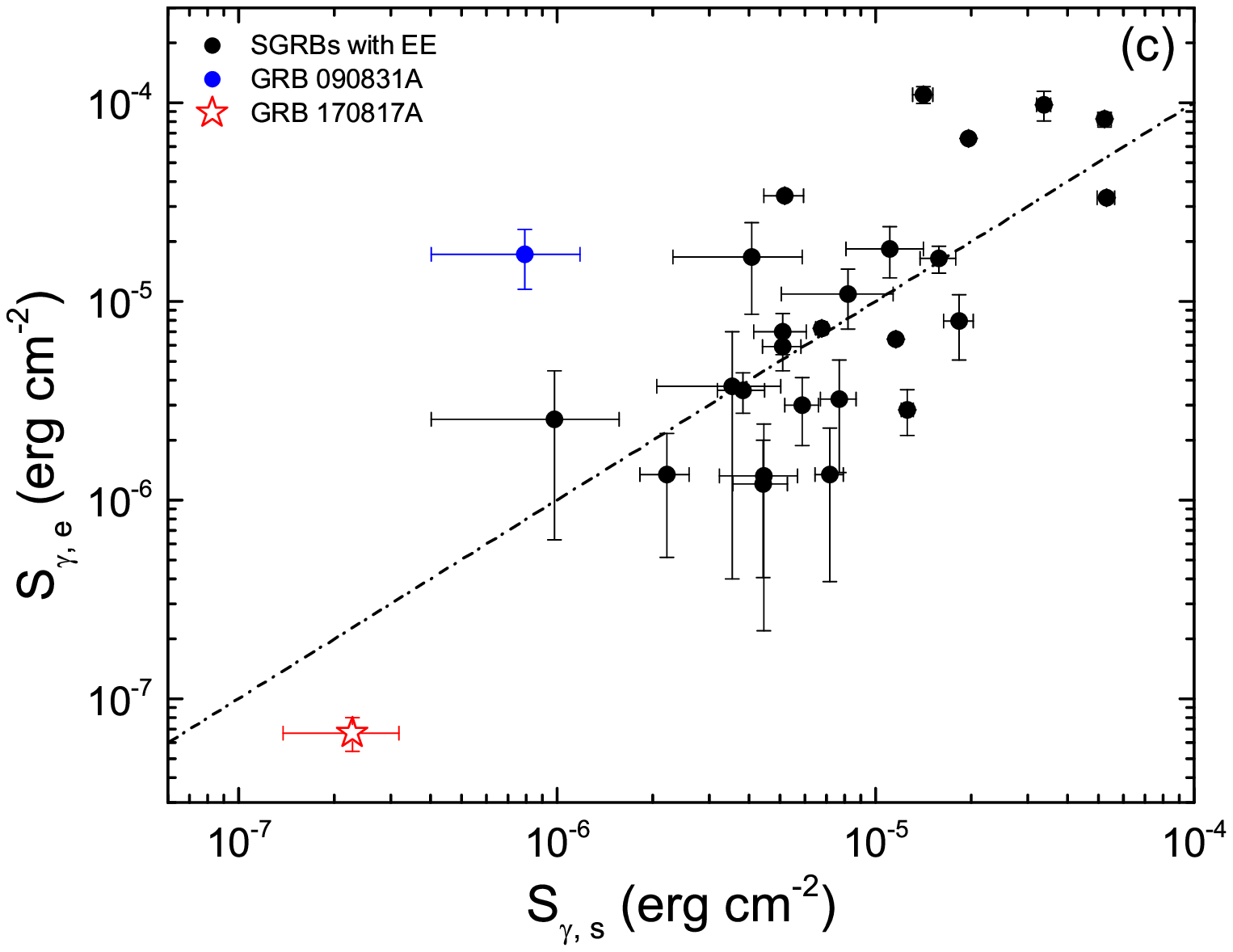}
\includegraphics[angle=0,scale=0.5]{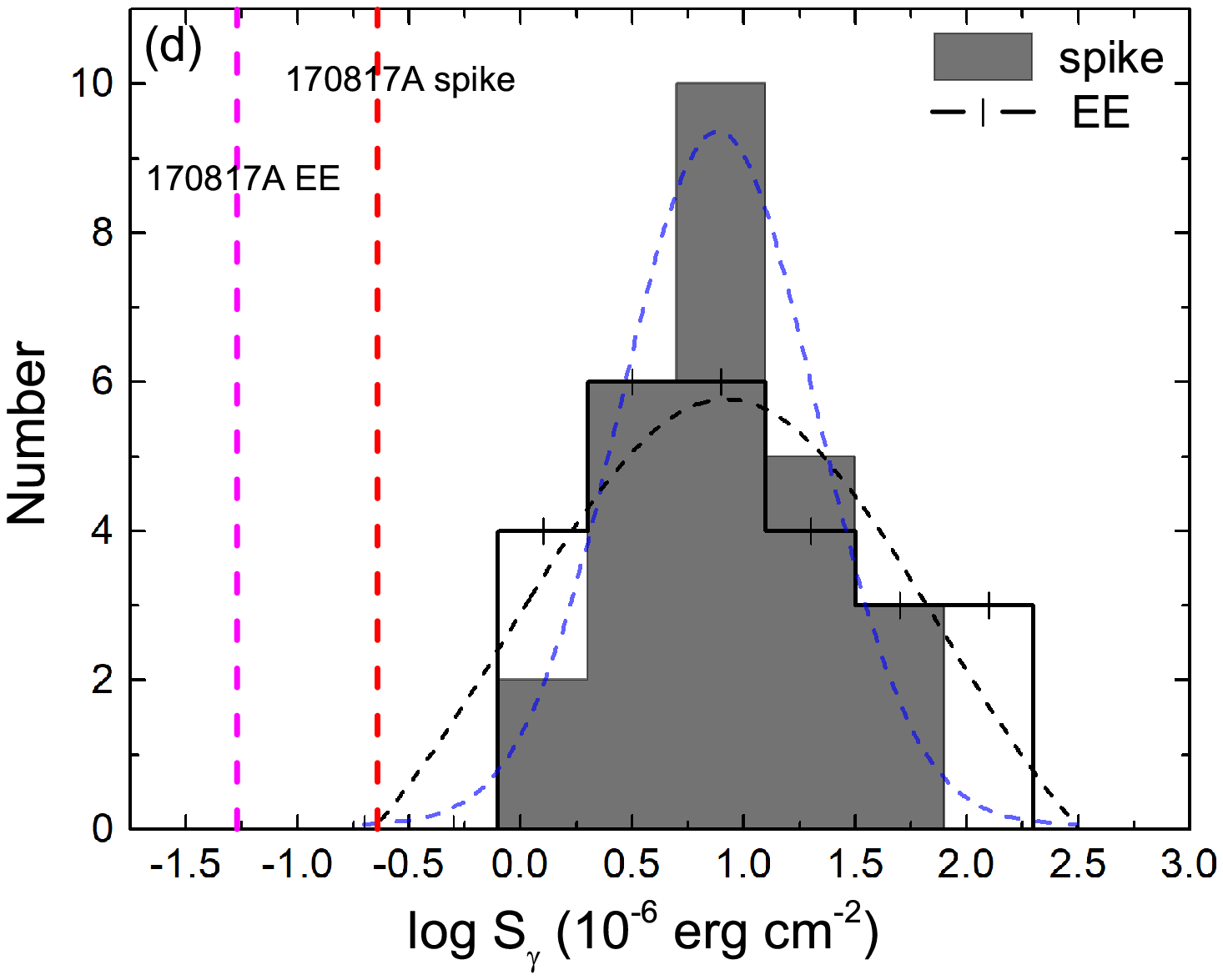}
\hfill\center \caption{$E_{\rm p, s}-E_{\rm p, e}$ (a) and $S_{\gamma, s}-S_{\gamma, e}$ (c)
correlations for the short GRBs with EE in our sample. The black dashed-dotted lines correspond to
$E_{\rm p, s}=E_{\rm p, e}$ in (a) and $S_{\gamma, s}=S_{\gamma, e}$ in (c), respectively. The
opened-red star is the GRB 170817A. (b) and (d) show the distributions of peak energy and fluence
for the hard spike and EE in our sample. Best-fit Gaussian profiles are denoted in black and blue
dashed lines, respectively. The red and magenta dashed vertical lines correspond to the hard spike
and EE of GRB 170817A, respectively.} \label{fig:Spectra}
\end{figure*}

Figure \ref{fig:Spectra}(a) shows the correlation of peak energy between the hard spike and
subsequent EE. We find that the peak energy of the hard spikes in our sample is higher than the
peak energy of the EE. This result is roughly consistent with results from the {\em Swift} and {\em
Fermi} eras (Norris \& Bonnell 2006; Kaneko et al. 2015). However for GRB 090831A, the peak energy
of its EE component is higher than that of its initial hard spike. In order to test whether the
peak energy value of the EE component of this case is valid, we compare the spectral fitting
models. We invoke the Bayesian information criterion (BIC)\footnote{BIC is a criterion for model
selection among a finite set of models. The model with the lowest BIC is preferred. BIC can be
expressed as: $BIC=\rm \chi^{2}+k\cdot \ln(n)$, where $k$ is the number of model parameters and $n$
is the number of data points. The strength of the evidence against the model with the higher BIC
value can be summarized as follows:\\(1) if $0<\Delta BIC<2$, the evidence against the model with
higher BIC is weak;\\ (2) if $2<\Delta BIC<6$, the evidence against the model with higher BIC is
positive;\\(3) if $6<\Delta BIC<10$, the evidence against the model with higher BIC is strong;\\(4)
if $10<\Delta BIC$, the evidence against the model with higher BIC is very strong.}, which is an
evaluation criterion for models defined by considering both the free parameters of the model and
the goodness of the fit (L\"{u} et al. 2017). We find the $\Delta$BIC $\gg10$, which means the CPL
model is strongly preferred, and the measured peak energy of EE component of GRB 090831A is valid.

Figure \ref{fig:Spectra}(b) presents the distribution of peak energy for the hard spike and
subsequent EE. The $E_{\rm p}$ distributions range from tens of keV to one thousand keV. Both
follow lognormal distributions with peaks at $E_{\rm p, s}=(447\pm78)$ keV and $E_{\rm p,
e}=(282\pm57)$ keV, respectively. We measure the difference of any pair of distributions with the
probability $P_{\rm KS}$ given by the Kolmogorov$-$Smirnov (KS) test, as proposed by Ashman et al.
(1994). The hypothesis that the two distributions are from the same parent sample is statistically
rejected if $P_{\rm KS}<10^{-4}$, and it is marginally rejected if $10^{-4}<P_{\rm KS}<0.1$. A
probability $P_{\rm KS}=1$ indicates that the two samples are identical. The KS test on our sample
returned a probability $P_{\rm KS}=0.11$, which indicates that the two peak energy distributions
are marginally similar, but are likely different. On the other hand, the minor differences between
the distributions may be not caused from the physically, but due to a selection effect.

Similarly, Figure \ref{fig:Spectra}(c) and (d) show the correlation and distribution of fluence for
the hard spike and subsequent EE. The fluence distributions of the hard spike and EE are also
lognormal with mean values of $\log S_{\rm \gamma, s}=(-5.11\pm 0.04) \rm~erg~cm^{-2}$ and $\log
S_{\rm \gamma, e}=(-5.08\pm 0.10)\rm ~erg~cm^{-2}$, respectively. The KS test on these two
distributions returned a probability $P_{\rm KS}=0.44$, which indicates that they cannot be
absolutely distinguished.

\begin{figure*}
\includegraphics[angle=0,scale=1.0]{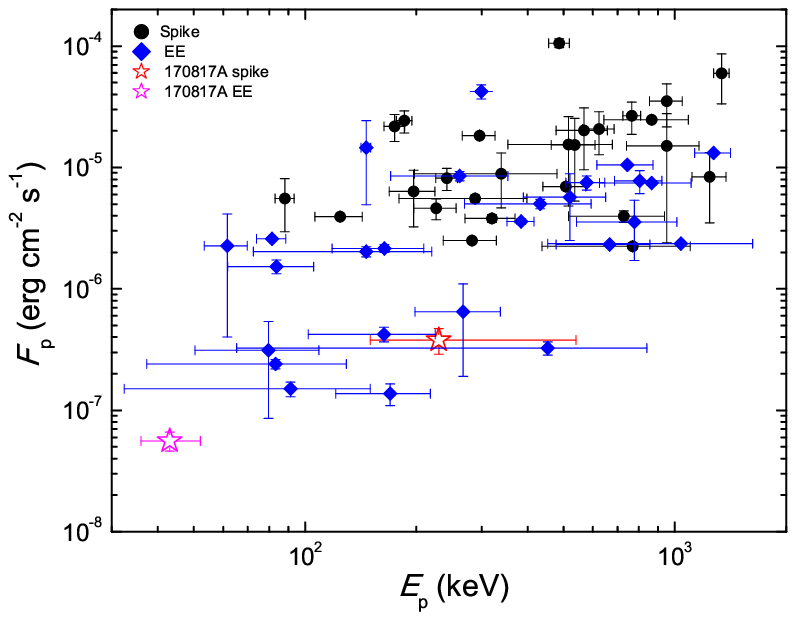}
\includegraphics[angle=0,scale=1.0]{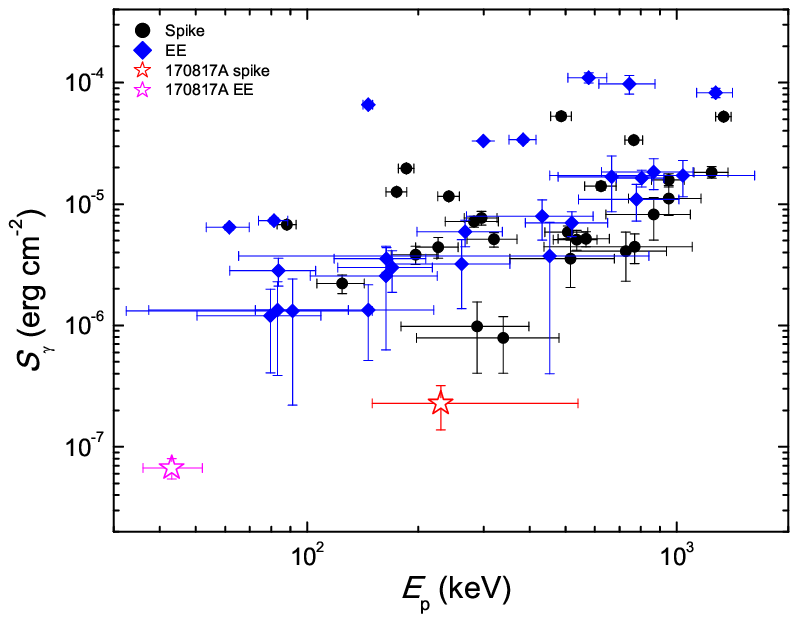}
\includegraphics[angle=0,scale=1.0]{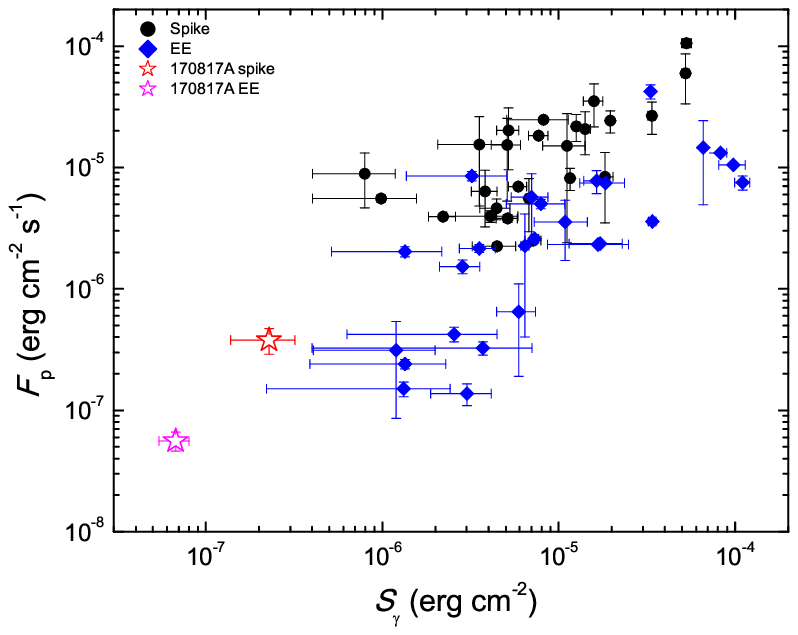}
\includegraphics[angle=0,scale=1.0]{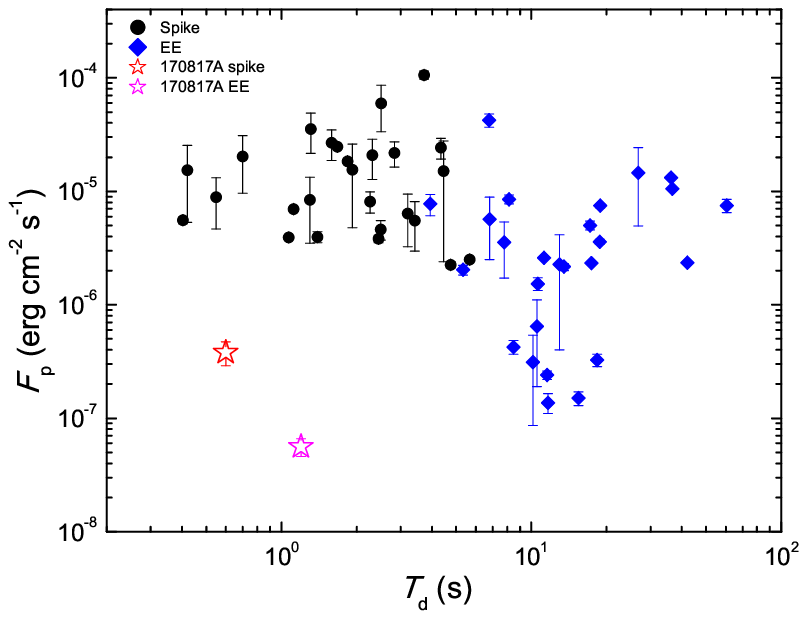}
\hfill\center \caption{$E_{\rm p}-F_{\rm p}$, $E_{\rm p}-S_{\rm \gamma}$, $S_{\gamma}-F_{\rm p}$,
and $F_{\rm p}-T_{\rm d}$ correlations of hard spike and EE in our sample. The red and magenta
opened stars correspond to the spike and EE of GRB 170817A, respectively.} \label{fig:Correlation}
\end{figure*}

In the CGRO era, a hardness-intensity correlation was discovered in the GRB prompt emission phase
in an analysis of BATSE data (Dezalay et al. 1998; Borgonovo \& Ryde 2001). Figure
\ref{fig:Correlation} presents the correlations of $E_{\rm p}-F_{\rm p}$, $E_{\rm p}-S_{\rm
\gamma}$, $S_{\gamma}-F_{\rm p}$, and $F_{\rm p}-T_{\rm d}$ of the hard spike and EE for the entire
sample. It seems to be that a higher peak flux or fluence generally has a higher peak energy, but
there is no significant correlation between the peak flux and the duration of the hard spike and
EE.

\begin{figure*}
\includegraphics[angle=0,scale=0.7]{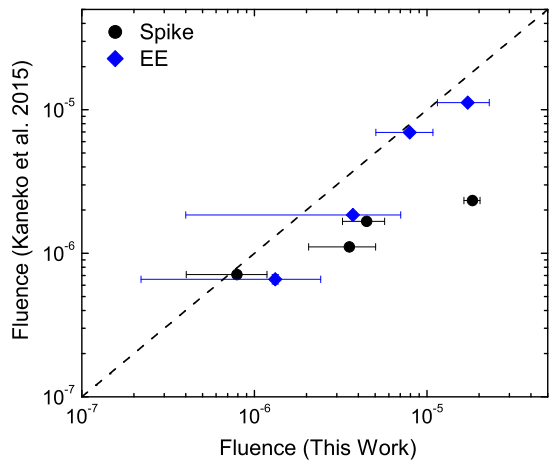}
\hfill\center \caption{Comparison of the estimated fluence in EE and hard spike with Kaneko et al.
(2015) for the four overlapping GRBs. The dashed line is the equivalent fluence between our
calculation and Kaneko et al. (2015).} \label{fig:compare}
\end{figure*}

One basic question is what is the difference between the estimated fluence in our calculation of EE
versus hard spike with that of Kaneko et al. (2015)? Figure \ref{fig:compare} shows the comparison
of the fluences between the two phases for four overlapping GRBs in the two samples. We find that
the fluence of both the hard spike and EE components in our calculation is larger than that of
Kaneko et al. (2015). Several factors may explain such a difference, e.g. the selection of
different energy bands used in calculating the fluence\footnote{The fluence in Kaneko et al. (2015)
is calculated in the energy range 15-350 keV, while we used the energy band 8-1000 keV to estimate
the fluence.}, the use of a non-standard definition of duration (5 seconds versus 2 seconds for
$T_{\rm 90}$), or different spectral fitting models.

\subsection{Short GRBs with EE vs. other typical short GRBs}
Troja et al. (2008) proposed that the short GRBs with EE and typical short GRBs may originate from
different progenitors (i.e., NS-BH or NS-NS mergers) by comparing the offsets from their host
galaxies. If this is the case, the different progenitors may correspond to different observational
properties. In this section, we compare the temporal and spectral properties between the short GRBs
with EE and other typical short GRBs in the {\em Fermi} era. Moreover, we also determine if there
is a difference between the hard spike and EE components in our sample compared to other typical
short GRBs.

\begin{figure*}
\begin{tabular}{cccccccccc}
\includegraphics[angle=0,scale=1.0]{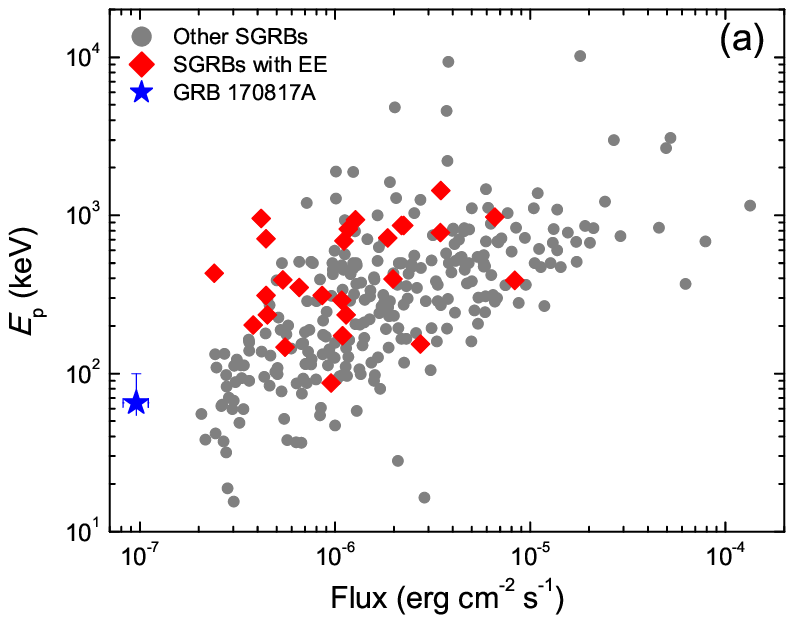}
\includegraphics[angle=0,scale=1.0]{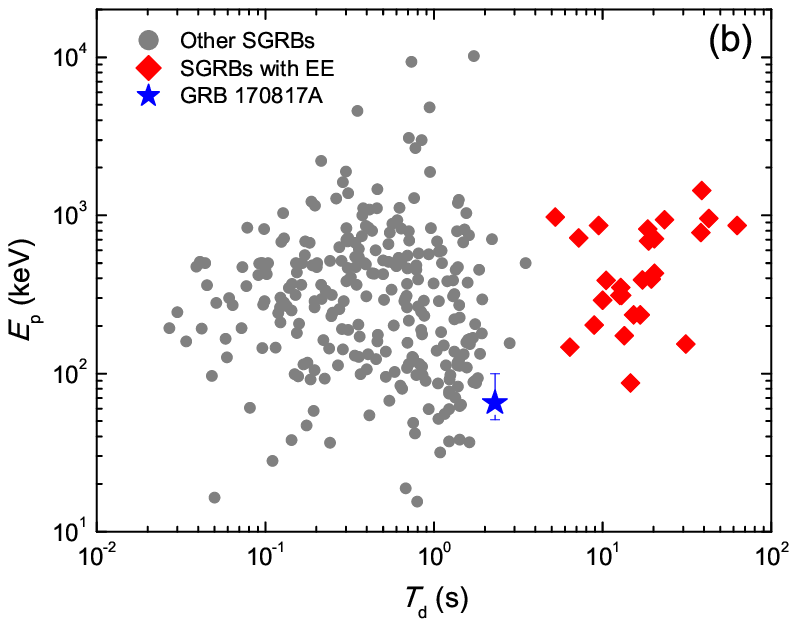}\\
\includegraphics[angle=0,scale=1.0]{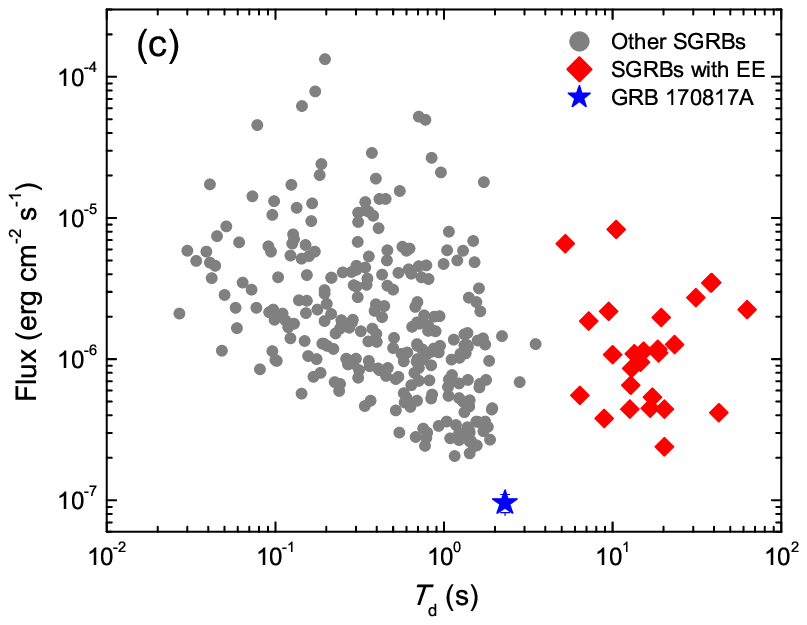}
\includegraphics[angle=0,scale=1.0]{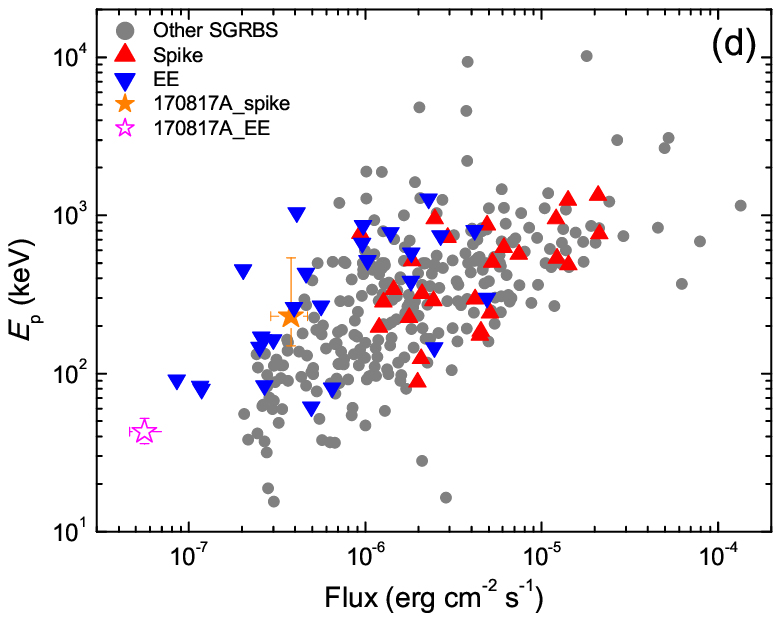}\\
\includegraphics[angle=0,scale=1.0]{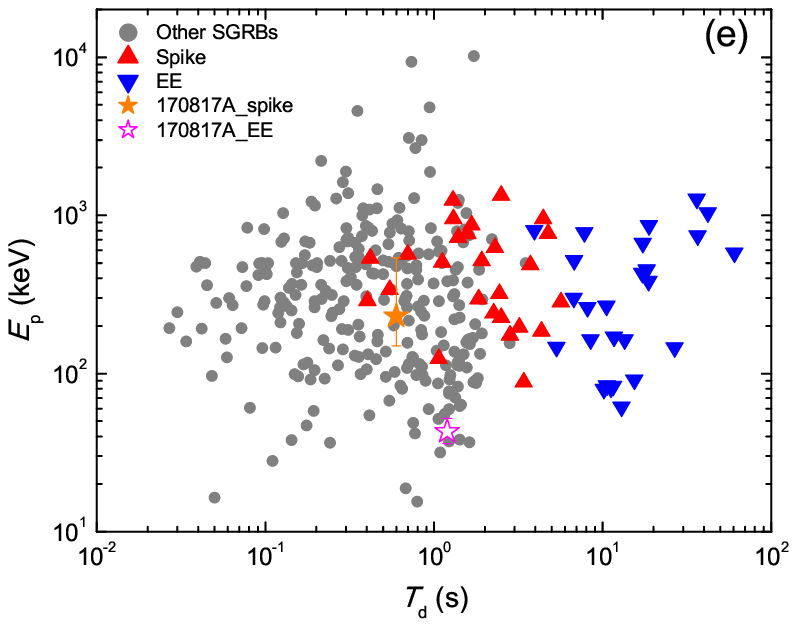}
\includegraphics[angle=0,scale=1.0]{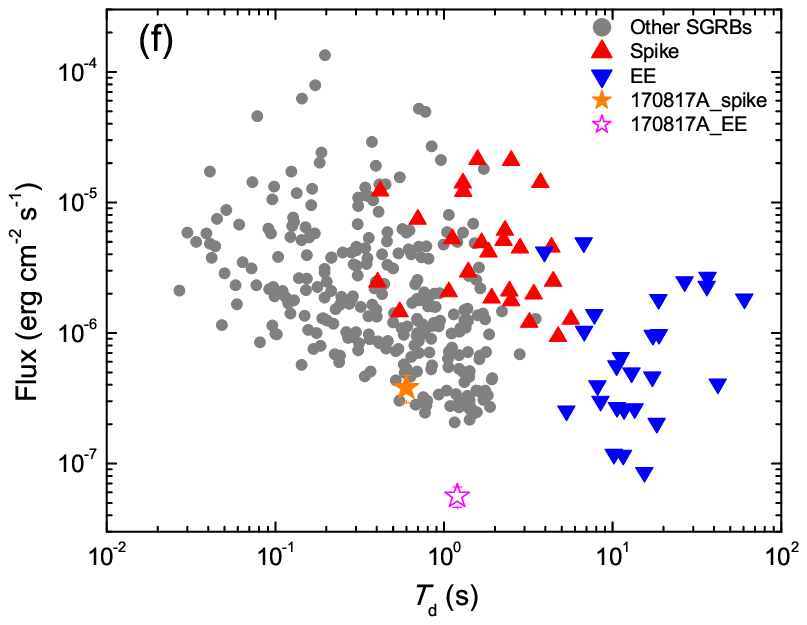}
\end{tabular}
\hfill\center \caption{(a)-(c): $E_{\rm p}-{\rm Flux}$, $E_{\rm p}-T_{\rm d}$, and ${\rm
Flux}-T_{\rm d}$ diagrams for short GRBs with EE in our sample (black diamonds) and other short
GRBs (gray dots; taken from Lu et al. 2017). (d)-(f): $E_{\rm p}-{\rm Flux}$, $E_{\rm p}-T_{\rm
d}$, and ${\rm Flux}-T_{\rm d}$ diagrams for hard spike (black triangles), EE (blue triangles), and
other short GRBs (gray dots). The hard spike and EE of GRB 170817A are marked with opened red and
magenta stars, respectively.} \label{fig:SGRBs}
\end{figure*}

Lu et al. (2017) presented a comprehensive analysis of short GRBs observed with {\em Fermi}/GBM and
derived a catalog of 275 typical short GRBs, which contains a peak energy distribution across a
wider range of tens to thousands of keV. We compared our sample to their more extensive catalog. In
Figure \ref{fig:SGRBs} we make some comparisons. The top three panels present the $E_{\rm p}-{\rm
Flux}$, $E_{\rm p}-T_{\rm d}$, and ${\rm Flux}-T_{\rm d}$ diagrams. The $E_{\rm p}$ is measured via
time-integrated spectral fits from the beginning of the spike to end of the EE, and ${\rm Flux}$
and $T_{\rm d}$ are the average flux and duration of the burst, respectively. Both $E_{\rm p}$ and
${\rm Flux}$ are not significantly different when comparing the short GRBs with EE with other
typical short GRBs.

\begin{figure*}
\includegraphics[angle=0,scale=0.7]{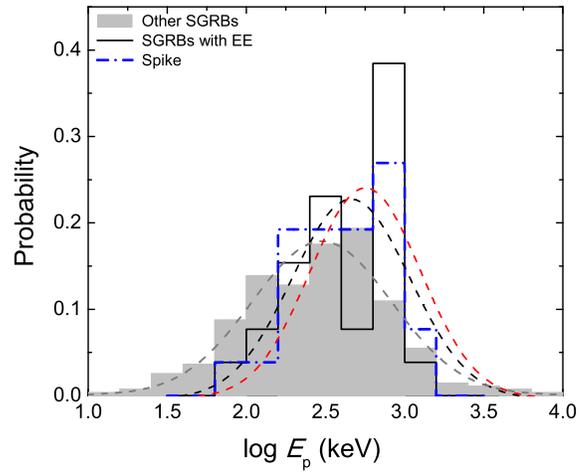}
\hfill\center \caption{Distributions of $E_{\rm p}$ for the time-integrated spectra of other short
GRBs (gray column), short GRBs with EE (black solid line), as well as the sample of hard spikes of
short GRBs with EE (blue dashed-dotted line). The best-fit Gaussian profiles correspond to the
respective colors, respectively.} \label{fig:Epcompare}
\end{figure*}

Moreover, it is necessary to test whether the hard spike and subsequent EE components are different
in comparison to other typical short GRBs. We separate the hard spike component and EE component to
measure their $E_{\rm p}$, ${\rm flux}$, and $T_{\rm d}$ values. The bottom three panels of Figure
\ref{fig:SGRBs} show the comparisons of $E_{\rm p}$, ${\rm flux}$, as well as $T_{\rm d}$ for the
hard spike, EE, and other short GRBs. The distribution of $E_{\rm p}$ for other typical short GRBs
is in a wider range. We find that the $E_{\rm p}$ of hard spike in our sample is tended to a higher
$E_{\rm p}$ side of other short GRBs, but EE component is tended to a lower $E_{\rm p}$ side of
other short GRBs. However, from the statistical point of view, the $E_{\rm p}$ distributions of
hard spike and EE components are not significant distinction with other typical short GRBs, this
may be caused by selection effects. Figure \ref{fig:Epcompare} presents the distributions of
$E_{\rm p}$ of SGRBs, SGRBs with EE (hard spike), and SGRBs with EE (time-integrated). We find that
they follow log-normal distributions with peaks at $E_{\rm p, s}=(302\pm 22)$ keV (SGRBs), $E_{\rm
p, s}=(380\pm 77)$ keV (SGRBs with EE), and $E_{\rm p, s}=(447\pm 78)$ keV (hard spikes). There are
minor differences between them, but nothing significant. From a statistical point of view, these
small differences may be due to the number of sources we used in our statistical analysis. Those
results suggest that one can not distinguish the progenitors of short GRBs with EE and other
typical short GRBs via their spectral properties alone.

\subsection{GRB 170817A in comparison to short GRBs with EE}
GRB 170817A is a short GRBs associated with the gravitational wave event (GW170817) from the double
NS merger which was recently detected by {\em Fermi}/GBM (Abbott et al. 2017; Goldstein et al.
2017). An analysis of the prompt emission of GRB 170817A shows that it consists of two different
components. The first component is a short hard spike whose spectrum can be fit by a CPL with peak
energy $E_{\rm p}=230^{+310}_{-80}$ keV. The preferred fit for the spectrum of the second component
is a blackbody model with $\rm kT=11.3^{+3.8}_{-2.4}$ keV (Goldstein et al. 2017; Zhang et al.
2018), but a non-thermal origin with a CPL model fit of $E_{\rm p}=43^{+9}_{-7}$ keV cannot be
ruled out (Zhang et al. 2018; Pozanenko et al. 2018). In this section, we compare the properties of
GRB 170817A to the short GRBs with EE in our sample.

Based on Figures \ref{fig:Spectra} and \ref{fig:Correlation}, we find that the peak energy and
fluence of the EE, and the fluence of the hard spike for GRB 170817A are the lowest in comparison
to all of our samples, but the peak energy of the hard spike for GRB 170817A is not the lowest.
Moreover, the peak flux of both the hard spike and EE of GRB 170817A are smallest out of all our
samples. The possible reason may be that GRB 170817A is an off-axis observation (Abbott et al.
2017; Alexander et al. 2018; Biehl et al.2018; Ioka \& Nakamura 2019).

\section{Conclusions}
We have presented a comprehensive temporal and spectral analysis for the GRB data observed with
{\em Fermi}/GBM during nine years of operation. By adopting the criterion of GRB 060614, we find
that a small fraction of GRBs observed by {\em Fermi}/GBM are similar to GRB 060614. The prompt
emission light curves of those events exhibit a hard spike initially and followed a soft tail
emission, so called short GRBs with extended emission. We try to determine the differences between
the initial hard spike and EE components of our sample, as well as the differences between short
GRBs with EE and other typical short GRBs observed by {\em Fermi}/GBM. Our results are summarized
as follows:
\begin{itemize}
\item We obtained a sample of 26 short GRBs with EE that were observed with {\em Fermi}/GBM.
    The peak energy of both the initial hard spike and subsequent EE component can be estimated
    via spectral fits with a CPL model or a Band model.
\item The peak energy of the EE components in our sample seems to be softer a little bit than
    the peak energy of the initial hard spike episodes, except for GRB 090831A, but the total
    fluence of the hard spike and subsequent EE are comparable with each other. Moreover, it
    seems to be that a higher peak flux or fluence generally has a higher peak energy for both
    the hard spikes and EE, but we do not find a significant correlation between peak flux and
    the duration of the hard spike and EE.
\item Both the peak energy and average flux of short GRBs with EE in our sample are not
    significantly different in comparison to other short GRBs observed by {\rm Fermi}/GBM.
    Moreover, the properties of the hard spike and followed EE components are also not
    significantly distinct in comparison with other short GRBs. These results suggest that the
    short GRBs with EE in our sample likely share a similar physical origin.
\end{itemize}

Moreover, the distribution of $E_{\rm p}$ for other typical short GRBs is in a wider range, and the
$E_{\rm p}$ of hard spike in our sample is tended to a higher $E_{\rm p}$ side of other short GRBs,
but EE component is tended to a lower $E_{\rm p}$ side of other short GRBs. However, from the
statistical point of view, the $E_{\rm p}$ distributions of hard spike and EE components are not
significant distinction with other typical short GRBs, this may be caused by selection effects and
sample selection criteria. On the other hand, the distribution of $E_{\rm p}$ of SGRBs with EE
(time-integrated) is between the that of SGRBs with EE (hard spike) and typical short SGRBs. The
minor differences between them seem to not be from the physically, but likely to be due to
selection effects. Those results also suggest that one can not distinguish the progenitors of short
GRBs with EE and other typical short GRBs via their spectral properties alone.

\section{Acknowledgements}
We acknowledge the use of the public data from the {\em Fermi} data archive. This work is supported
by the National Natural Science Foundation of China (Grant Nos.11922301, 11603006, 11851304 and
11533003), the Guangxi Science Foundation (grant No. 2017GXNSFFA198008, AD17129006, and
2018GXNSFGA281007). The One-Hundred-Talents Program of Guangxi colleges, Bagui Young Scholars
Program (LHJ), and special funding for Guangxi distinguished professors (Bagui Yingcai \& Bagui
Xuezhe).


\begin{table*}
\center \setlength{\tabcolsep}{0.11in} \caption{The spectral fitting parameters for our sample. }
\begin{tabular}{cccccccccccccc}
\hline
Trigger ID  &  Components & Model  & $T_{\rm d}^{a}$ & $E_{\rm p}^{b}$ & $\alpha^{c}$ & $S_{\gamma}^{d}$
& $F_{\rm p}^{e}$ & $(\chi^{2}/$dof) \\
&             &        & (s)             & (keV)           &              & ($\rm {erg~cm^{-2}}$) & ($\rm
{erg~cm^{-2}~s^{-1}}$)  &     \\
\hline
bn080807993	&	Spike	&	CPL	&	1.92 	&	515$\pm$162	 &	0.73$\pm$0.12	&	(3.55$\pm$1.49)e-6	&	
(1.55$\pm$1.07)e-5	&	234/240	\\
	        &	EE	    &	CPL	&	18.30 	&	453$\pm$387	 &	1.16$\pm$0.19	&	(3.72$\pm$3.32)e-6	&	
(3.25$\pm$0.40)e-7	&	183/241	\\
bn081110601	&	Spike	&	CPL	&	1.12 	&	507$\pm$67	 &	0.75$\pm$0.06	&	(5.89$\pm$0.71)e-6	&	
(6.95$\pm$0.07)e-6	&	256/237	\\
	        &	EE	    &	CPL	&	11.68 	&	170$\pm$49	 &	0.88$\pm$0.15	&	(3.01$\pm$1.13)e-6	&	
(1.37$\pm$0.28)e-7	&	246/239	\\
bn081129161	&	Spike	&	CPL	&	2.45 	&	320$\pm$49	 &	0.82$\pm$0.07	&	(5.12$\pm$0.70)e-6	&	
(3.81$\pm$0.28)e-6	&	242/237	\\
	        &	EE	    &	CPL	&	10.55 	&	267$\pm$69	 &	1.03$\pm$0.11	&	(5.92$\pm$1.47)e-6	&	
(6.47$\pm$4.56)e-7	&	254/237	\\
bn081215784	&	Spike	&	Band&	3.72 	&	487$\pm$32	 &	-0.53$\pm$0.03	&	(5.30$\pm$0.34)e-5	&	
(1.06$\pm$0.09)e-4	&	266/235	\\
	        &	EE	    &	CPL	&	6.77 	&	300$\pm$12	 &	0.77$\pm$0.02	&	(3.31$\pm$0.08)e-5	&	
(4.24$\pm$0.56)e-5	&	266/235	\\
bn090720710	&	Spike	&	CPL	&	0.42 	&	535$\pm$73	 &	0.37$\pm$0.08	&	(5.11$\pm$0.96)e-6	&	
(1.53$\pm$1.00)e-5	&	250/240	\\
	        &	EE	    &	CPL	&	6.81 	&	520$\pm$131	 &	0.87$\pm$0.09	&	(7.02$\pm$1.64)e-6	&	
(5.68$\pm$3.18)e-6	&	263/240	\\
bn090831317	&	Spike	&	CPL	&	0.55 	&	339$\pm$141	 &	1.10$\pm$0.16	&	(7.92$\pm$3.90)e-7	&	
(8.92$\pm$4.26)e-6	&	253/240	\\
	        &	EE	    &	CPL	&	42.20 	&	1037$\pm$584 &	1.51$\pm$0.07	&	(1.72$\pm$0.57)e-5	&	
(2.35$\pm$0.05)e-6	&	194/240	\\
bn090929190	&	Spike	&	CPL	&	1.84 	&	296$\pm$30	 &	0.32$\pm$0.07	&	(7.70$\pm$0.98)e-6	&	
(1.83$\pm$0.03)e-5	&	225/239	\\
	        &	EE	    &	CPL	&	8.15 	&	262$\pm$92	 &	0.70$\pm$0.18	&	(3.22$\pm$1.85)e-6	&	
(8.51$\pm$0.74)e-6	&	229/239	\\
bn091127976	&	Spike	&	CPL	&	2.26 	&	242$\pm$16	 &	1.36$\pm$0.03	&	(1.16$\pm$0.03)e-5	&	
(8.15$\pm$1.70)e-6	&	260/238	\\
	        &	EE	    &	CPL	&	12.97 	&	62$\pm$8	 &	1.75$\pm$0.07	&	(6.43$\pm$0.33)e-6	&	
(2.27$\pm$1.87)e-6	&	268/238	\\
bn100829876	&	Spike	&	CPL	&	2.83 	&	175$\pm$11	 &	0.68$\pm$0.04	&	(1.26$\pm$0.06)e-5	&	
(2.18$\pm$0.55)e-5	&	222/238	\\
	        &	EE	    &	CPL	&	10.61 	&	84$\pm$22	 &	1.14$\pm$0.15	&	(2.84$\pm$0.74)e-6	&	
(1.53$\pm$0.20)e-6	&	262/239	\\
bn110824009	&	Spike	&	CPL	&	1.30 	&	1243$\pm$132 &	0.68$\pm$0.04	&	(1.83$\pm$0.20)e-5	&	
(8.41$\pm$4.92)e-6	&	243/238	\\
	        &	EE	    &	CPL	&	17.20 	&	432$\pm$162	 &	1.02$\pm$0.11	&	(7.93$\pm$2.88)e-6	&	
(5.02$\pm$0.45)e-6	&	223/239	\\
bn111012811	&	Spike	&	CPL	&	1.07 	&	124$\pm$18	 &	0.48$\pm$0.10	&	(2.21$\pm$0.39)e-6	&	
(3.93$\pm$0.13)e-6	&	288/237	\\
	        &	EE	    &	CPL	&	5.33 	&	146$\pm$74	 &	0.99$\pm$0.22	&	(1.34$\pm$0.83)e-6	&	
(2.03$\pm$0.20)e-6	&	192/239	\\
bn120119229	&	Spike	&	CPL	&	1.40 	&	727$\pm$212	 &	0.58$\pm$0.13	&	(4.09$\pm$1.78)e-6	&	
(3.97$\pm$0.44)e-6	&	251/239	\\
	        &	EE	    &	CPL	&	17.36 	&	666$\pm$189	 &	0.51$\pm$0.14	&	(1.67$\pm$0.81)e-5	&	
(2.32$\pm$0.03)e-6	&	200/239	\\
bn120304248	&	Spike	&	CPL	&	1.31 	&	951$\pm$96	 &	0.48$\pm$0.05	&	(1.58$\pm$0.20)e-5	&	
(3.53$\pm$1.36)e-5	&	271/239	\\
	        &	EE	    &	CPL	&	3.94 	&	804$\pm$118	 &	0.62$\pm$0.06	&	(1.64$\pm$0.26)e-5	&	
(7.75$\pm$1.66)e-6	&	275/239	\\
bn120605453	&	Spike	&	CPL	&	4.75 	&	769$\pm$332	 &	1.26$\pm$0.08	&	(4.46$\pm$1.22)e-6	&	
(2.25$\pm$0.14)e-6	&	266/235	\\
	        &	EE	    &	CPL	&	15.48 	&	91$\pm$59	 &	1.46$\pm$0.32	&	(1.32$\pm$1.10)e-6	&	
(1.50$\pm$0.21)e-7	&	245/237	\\
bn130628531	&	Spike	&	CPL	&	4.65 	&	283$\pm$46   &	1.09$\pm$0.06	&	(7.18$\pm$0.74)e-6	&	
(2.51$\pm$0.10)e-6	&	243/239	\\
	        &	EE	    &	CPL	&	11.56 	&	83$\pm$46	 &	1.49$\pm$0.28	&	(1.34$\pm$0.95)e-6	&	
(2.40$\pm$0.21)e-7	&	238/239	\\
bn131108862	&	Spike	&	CPL	&	0.70 	&	568$\pm$90	 &	0.82$\pm$0.06	&	(5.19$\pm$0.74)e-6	&	
(2.03$\pm$1.07)e-5	&	271/241	\\
	        &	EE	    &	CPL	&	18.76 	&	384$\pm$32	 &	0.96$\pm$0.03	&	(3.39$\pm$0.16)e-5	&	
(3.60$\pm$0.18)e-6	&	242/241	\\
bn140308710	&	Spike	&	CPL	&	3.20 	&	197$\pm$28	 &	0.60$\pm$0.09	&	(3.83$\pm$0.64)e-6	&	
(6.36$\pm$3.12)e-6	&	263/242	\\
	        &	EE	    &	CPL	&	13.50 	&	164$\pm$45	 &	1.19$\pm$0.12	&	(3.55$\pm$0.82)e-6	&	
(2.16$\pm$0.16)e-6	&	276/242	\\
bn141229492	&	Spike	&	CPL	&	2.50 	&	226$\pm$30	 &	0.38$\pm$0.10	&	(4.43$\pm$0.86)e-6	&	
(4.62$\pm$0.91)e-6	&	299/237	\\
	        &	EE	    &	CPL	&	10.17 	&	80$\pm$29	 &	0.88$\pm$0.26	&	(1.20$\pm$0.79)e-6	&	
(3.12$\pm$2.26)e-7	&	243/237	\\
bn150127398	&	Spike	&	CPL	&	1.68 	&	866$\pm$223	 &	0.51$\pm$0.12	&	(8.20$\pm$3.14)e-6	&	
(2.47$\pm$0.03)e-5	&	249/240	\\
	        &	EE	    &	CPL	&	7.80 	&	778$\pm$236	 &	0.76$\pm$0.10	&	(1.09$\pm$0.36)e-5	&	
(3.55$\pm$1.84)e-6	&	224/240	\\
bn150510139	&	Spike	&	CPL	&	1.58 	&	765$\pm$43	 &	0.54$\pm$0.03	&	(3.37$\pm$0.17)e-5	&	
(2.67$\pm$0.79)e-5	&	251/239	\\
	        &	EE	    &	CPL	&	36.62 	&	744$\pm$129	 &	0.75$\pm$0.06	&	(9.76$\pm$1.68)e-5	&	
(1.05$\pm$0.01)e-5	&	222/239	\\
bn150702998	&	Spike	&	CPL	&	4.46 	&	951$\pm$211	 &	0.64$\pm$0.09	&	(1.11$\pm$0.30)e-5	&	
(1.51$\pm$1.27)e-5	&	268/239	\\
	        &	EE	    &	CPL	&	18.90 	&	865$\pm$240	 &	0.86$\pm$0.09	&	(1.84$\pm$0.53)e-5	&	
(7.46$\pm$0.33)e-6	&	231/239	\\
bn160721806	&	Spike	&	CPL	&	0.40 	&	288$\pm$109	 &	0.30$\pm$0.32	&	(9.83$\pm$5.81)e-7	&	
(5.56$\pm$0.39)e-6	&	241/238	\\
	        &	EE	    &	CPL	&	8.49 	&	163$\pm$61	 &	0.88$\pm$0.23	&	(2.55$\pm$1.92)e-6	&	
(4.24$\pm$0.59)e-7	&	215/238	\\
bn161218356	&	Spike	&	CPL	&	4.35 	&	185$\pm$9	 &	0.73$\pm$0.03	&	(1.96$\pm$0.06)e-5	&	
(2.43$\pm$0.50)e-5	&	258/239	\\
	        &	EE	    &	CPL	&	26.76 	&	146$\pm$4	 &	0.46$\pm$0.02	&	(6.60$\pm$0.13)e-5	&	
(1.46$\pm$0.97)e-5	&	226/239	\\
bn170115743	&	Spike	&	CPL	&	2.51 	&	1337$\pm$65	 &	0.55$\pm$0.03	&	(5.23$\pm$0.26)e-5	&	
(5.98$\pm$2.63)e-5	&	247/239	\\
	        &	EE	    &	CPL	&	36.28 	&	1272$\pm$142 &	0.97$\pm$0.03	&	(8.26$\pm$0.70)e-5	&	
(1.32$\pm$0.01)e-5	&	246/239	\\
bn170527480	&	Spike	&	CPL	&	2.31 	&	624$\pm$61	 &	0.71$\pm$0.04	&	(1.41$\pm$0.11)e-5	&	
(2.08$\pm$0.81)e-5	&	273/239	\\
	        &	EE	    &	CPL	&	60.48 	&	577$\pm$69	 &	0.79$\pm$0.06	&	(1.10$\pm$0.11)e-4	&	
(7.50$\pm$0.99)e-6	&	211/239	\\
bn170626401	&	Spike	&	CPL	&	3.42 	&	88$\pm$5     &	0.59$\pm$0.05	&	(6.77$\pm$0.27)e-6	&	
(5.54$\pm$2.58)e-6	&	246/239	\\
	        &	EE	    &	CPL	&	11.24 	&	81$\pm$7	 &	0.92$\pm$0.06	&	(7.28$\pm$0.40)e-6	&	
(2.59$\pm$0.08)e-6	&	252/241	\\

\hline
\end{tabular}
\footnotesize{ \begin{flushleft}
$\rm{^{a}}$ Durations of the hard spike and EE for our sample.\\
$\rm{^{b}}$ Peak energy of CPL or Band model fits of the hard spike and EE.\\
$\rm{^{c}}$ The low-energy photon index of CPL or Band model fits for the hard spike and EE.\\
$\rm{^{d}}$ Energy fluence of the hard spike and EE.\\
$\rm{^{e}}$ Peak flux of CPL or Band model fits of the hard spike and EE.\\
\end{flushleft}}
\label{specanlysis}
\end{table*}

\end{document}